\documentclass[12pt,letterpaper]{article}
\usepackage{amsmath,amssymb} 
\title{The Relativistic Field Theory of Fluids}
\author{Sylvan A. Jacques}
\begin{document}
\maketitle
\tableofcontents
\begin{abstract}
Classical relativistic field theory is applied to perfect and magneto-hydrodynamic flows. The fields for Hamilton's principle are shown to be the Lagrangian coordinates of the fluid elements, which are potentials for the matter current 4-vector and the electromagnetic field 2-form. The energy momentum tensor and equations of motion are derived from the fields. In this way the theory of continua is shown to have the same form as other field theories, such as electromagnetism and general relativity.

Waves are treated as an example of the power of field theoretic methods. The average or background flow and the waves are considered as two interacting components of the system. The wave-background interaction involves the transfer of energy and momentum between the waves and the average flow, but the total energy and momentum are conserved. The average Lagrangian for the total wave-background system is found by expanding the Lagrangian about the background flow and averaging over the phase. The total energy-momentum tensor is constructed, and the conservation of energy and momentum are discussed. Varying the wave amplitude gives the dispersion and polarization relations for the waves, and varying the phase gives the rays and conservation of wave quanta (or wave action). The wave quanta move with the group velocity along the bi-characteristic rays.
\end{abstract}
\part{Field Theory for Fluids}
\section{Introduction}
The theory is both Lorentz and generally covariant on space time M, so the physical and geometric quantities are independent of the coordinate system, i.e. they are tensor fields. We use coordinate free notation, and units so that the speed of light = 1. The natural mathematical framework is differential geometry \cite{AMP, mtw, thir}. 

The theory is introduced in Part 1 and  is applied to waves in Part 2. Relativistic thermodynamics is discussed in \S 2. In \S 3 the Lagrangian coordinates are shown to be the three scalar fields for Hamilton's principle. They are also the potentials for the matter current 4-vector and the electromagnetic field 2-form in magnetohydrodynamics (MHD) and plasmas. In \S 4, 5, and 6, the energy momentum tensor and the conservation equations for rest mass (or the number of particles) and for the energy and momentum, are discussed for perfect fluids and MHD. In Part 2, the theory is applied to waves. The fields for the waves are their Lagrangian coordinates, which are the oscillation centers.
\section{Relativistic Thermodynamics of an Ideal Gas}
A \emph{proper} quantity is one that moves with the flow (i.e. is measured with respect to a comoving coordinate system) and so is coordinate independent. It is a scalar field, which is a rank 0 tensor field (tensors include scalars and vectors). I use coordinate independent notation to emphasize the fact that tensors are independent of the coordinate system. The following thermodynamic quantities are scalar fields since they are all \emph{proper} quantities.

\begin{itemize} 
\item $n' =$ proper number density
\item $m =$ rest mass of fluid particles
\item $n = n'm =$ proper density of rest mass
\item $ne =$ internal (thermal) energy density
\item $e =$ thermal energy per unit mass
\item $s =$ specific entropy
\item $ \rho = n \, + \, ne =$ total energy density
\item $T =$ temperature
\item $p = nT/m =$ isotropic pressure
\item $\gamma = c_{p}/c_{v} = \partial \ln p/\partial \ln n|_{s} =$ the ratio of specific heats
\item $f = (\rho  +  p)/n = 1  +  e  +  p/n =$ the index or relativistic specific enthalpy
\end{itemize}
If $n$ and $s$ are the independent thermodynamic variables, the first law of thermodynamics is the conservation of energy:
\begin{equation}
T ds = de  +  p \, d\left(\frac{1}{n}\right)
\end{equation}
If the flow is \emph{isentropic}, $ds = 0$ and $s$ is constant. Then the thermodynamic quantities are functions of $n$ alone, and equation (1) gives
\begin{align}
p &= n^2 \frac{de}{dn}  &  dp &= n \, df  &  \frac{d\rho}{dn} &= f\,.
\end{align}
\section{The Mathematical Formalism}
\subsection{Three Coordinate Systems}
Assume that there is no gravity so that spacetime $M$ is flat. Consider the following three coordinate systems for $M$:
\begin{enumerate}
\item The observer's global  orthonormal coordinate system $x = (x^0, x^1, x^2, x^3)$.
\item A local comoving coordinate system (CMCS) $y = (y^{0}, y^{1}, y^{2}, y^{3})$ at each point $x \in M$. This coordinate system is also orthonormal, with $u = \partial /\partial y^0$.
\item A global CMCS $z = (z^{0}, z^{1}, z^{2}, z^{3})$. $z$ is not orthonormal, so the metric $g$ is not diagonal. In this coordinate system, $u \parallel \partial/\partial z^0$, and $z^{0}$ is closely related to the proper time for a fluid element. The $z^{k}$ are the fields which label the worldlines. 
\end{enumerate}
\subsection{Differential p-Forms}
Let $\Lambda^{p}(M) (\Lambda_{p}(M))$ be the set of p-forms (p-vectors) on $M$. A contravariant vector field $v \in \Lambda_1(M)$ or a 1-form $\alpha \in \Lambda^{1}(M)$ can be written in terms of the dual coordinate basis $e_{\alpha} = \partial/\partial x^{\alpha}$ and $e^{\alpha} = \textup{d}x^{\alpha}$ as
\begin{align*}
v &= v^{\mu} e_{\mu}  &  \alpha &= \alpha_{\mu} e^{\mu} 
\end{align*}
The exterior derivative $\textup{d}\colon \Lambda^{p} \to \Lambda^{p+1}$ of $\alpha$ is $\textup{d}\alpha = \nabla \wedge \alpha$. A generalized inner product on forms and p-vectors can be defined as the adjoint of the exterior product:
\begin{gather*}
(\quad|\quad)_{p} \colon \Lambda^{q} \times \Lambda^{p} \to \Lambda^{q-p} \colon (\alpha,\beta) \to (\alpha|\beta)_{p} \equiv \alpha(\beta) \qquad q \ge p,\; \textup{defined by} \\
(\alpha(\beta)|\gamma) = \alpha(\beta,\gamma) \equiv (\alpha|\beta \wedge \gamma)  \quad \textup{where} \quad \alpha \in \Lambda^{q}, \; \beta \in \Lambda^{p}, \; \gamma \in \Lambda^{q-p}.
\end{gather*}
The Hodge operator $*$ is defined by $* \colon \Lambda^{p} \to \Lambda^{4-p} \colon \beta \to *\beta$
\begin{gather}
\alpha \wedge *\beta = (\alpha|\beta) \tau \; \Leftrightarrow \; *\beta = (\tau|\beta)_{p} = \tau(\beta), \quad \textup{where} \quad \alpha, \beta \in  \Lambda^{p},\\
\tau = *1 = \sqrt{-\det g} \, \epsilon, \quad \textup{and} \quad \epsilon = \textup{d}x^{0} \wedge \textup{d}x^{1} \wedge \textup{d}x^{2} \wedge \textup{d}x^{3}.\nonumber
\end{gather}
$\tau$ is the volume element 4-form, $\epsilon$ is the 4D Levi-Civita tensor, and $g$ is the metric. If $x$ is a orthonormal coordinate system, det$g = - 1$, and $\tau = \epsilon = \textup{d}^4x$.
\subsection{The Scalar Fields}
\subsubsection{Equation of continuity}
Compare the problem of a fluid in 2D spacetime $(t,x) = (x^0,x^1)$ with electromagnetism (EM). (The following is also true in 4D spacetime). The equation of continuity $dj = 0$ is analogous to Maxwell's homogeneous equation $dF = 0$. $j$ is the dual of the matter current vector $J = nu = *j$, and $j = *J = **j$. We have
\begin{equation} 
\nabla \cdot J = *(dj) = 0 \quad  \Rightarrow \quad  dj = 0 \quad \Rightarrow \quad j = dz
\end{equation}
Thus the equation of continuity implies the existence of the scalar field $z(t,x)$ such that $j = dz$, just as in EM where we have the vector field $A$ and $F = dA$. In both cases we use the fact that a closed form is exact. In 4D spacetime, there are 3 scalar fields, and
\begin{equation*}
dj = 0 \quad \Rightarrow \quad j = \textup{d}z^1 \wedge \textup{d}z^2 \wedge \textup{d}z^3
\end{equation*}
Often a density r is introduced so that $z^i$ can be a Cartesian coordinate system, and then $j = r \, dz^1 \wedge dz^2 \wedge dz^3$. The invariant density is
\begin{equation*}
n = \sqrt{(j|j)}
\end{equation*}

Consider a fluid in flat spacetime $M \subseteq \mathbb{R}^{4}$  (no gravity).  We need to express the Lagrangian in terms of the fields and their first partial derivatives. For the initial value problem, $x^{0} = t \ge 0$. Define the initial configuration on the observer's 3D hypersurface $\Sigma \subseteq \mathbb{R}^{3}$ defined by $x^{0} = t = 0$. The equations of motion determine the flow on $M$ for $t > 0$. 

The three scalar fields are the Lagrangian coordinates $Z \equiv (z^{1}, z^{2}, z^{3})$. They are a coordinate system on $\Sigma$ and move with the fluid, and are fundamental to the theory of all continua. The flow is determined by the projection Z of a fluid element at $x \in M$ onto its initial position $Z(x)$ on $\Sigma$:
\begin{gather*}
Z \colon M \to \Sigma \colon x \to Z(x) = (z^1(x), z^2(x), z^3(x)) \quad \textup{where} \\
z^{k}(0,x^{i}) = x^{k} \quad \textup{and} \quad dz^{k}(u) = u^{\alpha} \frac{\partial z^{k}}{\partial x^{\alpha}} = 0 
\end{gather*}
and the unit 4-velocity u is tangent to the worldlines of the fluid elements (i.e., the streamlines). The $z^k$ have the following properties:
\begin{enumerate}
\item They are \emph{potentials} for the matter current 4-vector $J = nu$ and the 2-form of the electromagnetic field $F$.
\item They are scalar fields which label fluid elements and so are constant along worldlines ($dz^k(u) = 0$).
\item They label worldlines, so varying the $z^{k}$ varies the worldlines of the fluid elements.
\item By using the fields $z^k$ we automatically satisfy the equation of continuity $dj = 0$ and conserve matter. 
\end{enumerate}
\subsection{The Matter Current and Proper Number Density}
Let $n_{0} \colon \Sigma \to \mathbb{R} \colon z^{k} \to n_{0}(z^{k})$ be the number density on the initial hypersurface $\Sigma$. The 3-form of a fluid element on the $\Sigma$ is
\begin{equation*}
j_0 = n_0 \,\textup{d}z^1 \wedge \textup{d}z^2 \wedge \textup{d}z^3 \quad  \textup{and} \quad N = \int_{\Sigma}{j_0}
\end{equation*}
is the total number of particles (or the total rest mass).  The flow is described by the matter current 4-vector $J = nu \in \Lambda_{1}(M)$, or its dual 3-form $j = *J \in \Lambda^{3}(M)$. $j = Z^{*}(j_{0})$ is the pull-back of $j_{0}$ from $\Sigma$ to $M$ by $Z$.
\begin{gather*}
Z^* \colon \Lambda^3(\Sigma) \to \Lambda^3(M) \colon j_0 \to Z^*(j_0) = \\ j =  r \: \textup{d}z^1 \wedge \textup{d}z^2 \wedge \textup{d}z^3 
= \frac{r}{3!} \frac{\partial (z^1, z^2, z^3)}{\partial (x^{\alpha}, x^{\beta}, x^{\gamma})} \, e^{\alpha \beta \gamma} = j_{\alpha \beta \gamma}  e^{\alpha \beta \gamma}/3!
\end{gather*}
where $e^{\alpha \beta \gamma} = e^{\alpha} \wedge e^{\beta} \wedge e^{\gamma}$ and $r = n_0 \circ Z$.

The matter current 4-vector is
\begin{equation}
J = *j = \epsilon(j) = \epsilon^{\alpha \beta \gamma \delta} j_{\alpha \beta \gamma} e_{delta}/3! = J^{\delta} e_{\delta} .
\end{equation}
Thus the components of $J$ and $j$ are
\begin{align*}
J^0 = j_{123} &= r \frac{\partial (z^1, z^2, z^3)}{\partial (x^{1}, x^{2}, x^{3})} = \gamma n & J^1 = - j_{023} &= - r \frac{\partial (z^1, z^2, z^3)}{\partial (x^{0}, x^{2}, x^{3})} = \gamma n v^1\\ 
J^{2} = j_{013} &= r \frac{\partial (z^1, z^2, z^3)}{\partial (x^{0}, x^{1}, x^{3})} = \gamma n v^{2} & J^{3} = - j_{012} &= - r \frac{\partial (z^1, z^2, z^3)}{\partial (x^{0}, x^{1}, x^{2})} = \gamma n v^{3} 
\end{align*}
These expressions show that the fields $z^{i}$ are indeed the potentials for the components of $J$. They also give the 3-velocity $v^{i} = J^{i}/J^{0}$ in terms of the $z^{i}$, and the components of the 4-velocity $u$; $u^{0} = \gamma = 1/sqrt{(1 - v^{2})}$ and $u^{i} = \gamma v^{i}$.

The proper number density n is
\begin{gather} \label{eq:nn}
n^2 = (j|j) = (*J|*J) = r^2 (\textup{d}z^{1} \wedge \textup{d}z^{2} \wedge \textup{d}z^{3}|\textup{d}z^{1} \wedge \textup{d}z^{2} \wedge \textup{d}z^{3})\\
= r^{2} \textup{det} (\textup{d}z^{i}|\textup{d}z^{j}) \nonumber
\end{gather}
$r$ is constant along streamlines since the equation of continuity is
\begin{equation}
\textup{d}j = \textup{d}(*J) = *(\nabla \cdot J) = \textup{d}r \wedge \textup{d}^{3}z = 0  \Rightarrow  \textup{d}r(u) = 0  \Rightarrow  r = r(z^{i})
\end{equation}
The equation of continuity follows from d$j_{0} = 0$, since there are no 4-forms in $\Sigma$:
\begin{equation*}
\textup{d}j = \textup{d} (Z^{*}j_{0}) = Z^{*}(\textup{d}j_{0}) = 0, \quad \textup{since} \quad \textup{d}\circ Z^{*} = Z^{*}\!\circ \, \textup{d}
\end{equation*}
A current $J$ with $\nabla \cdot J = 0$ is said to be conserved. Associated with any conserved current is a conserved 'charge', which is the total number of particles (or total rest mass)
\begin{equation*}
N = \int_{\Sigma}{j} = \int_{\Sigma}{j_{0}} \int_{\Sigma}{J^{0}} \textup{d}^{3}x
\end{equation*}
\subsection{Decomposing Vectors and Tensors}
The metric $g$ and 4-velocity $u$ define a spatial projection operator $P = g + uu$ such that $P(u) = 0, P = P^{T}, P^{2} = P, Tr(P) = 3$, and $P = \textup{diag}(0, 1, 1, 1)$ in a local CMCS. $P$ projects a vector field onto the 3D space orthogonal to $u$ at any $x \in M$. A vector field A can be written as the sum of a vector parallel to $u$ and a vector orthogonal to $u$:
\begin{equation*}
A = - A(u) u + P(A) = au + A'.
\end{equation*}
$a$ is the temporal part of $A$ and $A'$ is the spatial part. $a$ and $A'$ are proper, coordinate independent quantities. $a$ is a scalar field and $A'$ is a spatial vector field.

If $T$ is a second order tensor field, decompose it as follows;
\begin{equation*}
T = T(u,u) - u P(T(u)) - P(T^{\dag}(u)) u + P \cdot T \cdot P
\end{equation*}
If T is the energy-momentum tensor,
\begin{itemize}
\item $T(u,u) =$ proper energy density
\item $- P(T(u)) = $ proper energy flux
\item $- P(T^{T}(u)) = $ proper momentum density, and
\item $P T P = $ proper stress tensor
\end{itemize}
Only proper quantities appear in the equations of motion, but the observer measures quantities with respect to his coordinate system. Observed and proper quantities are related by a Lorentz transformation with the 3-velocity $v(x)$ of the fluid in the observer's $x$ coordinate system.

\section{Energy-Momentum Tensor and\\Conservation Equations}

We must write the Lagrangian $L$ in terms of the fields $z^{k}$ and their first partial derivatives d$z^{k}$, and calculate the energy-momentum tensor (EMT)
\begin{equation} \label{eq:defT}
T = Lg - DL \quad \textup{where} \quad D =  \textup{d}z^{k} \otimes \frac{\partial}{\partial \textup{d}z^{k}} = D_{\mu}^{\nu} e^{\mu} \otimes e_{\nu}  \; ; \; D_{\mu}^{\nu} = z^{k}_{\mu} \frac {\partial}{\partial z^{k}_{\nu}}
\end{equation}
The fields are scalars so $T = T^{T}$ and $DL$ are symmetric. $DL$ is spatial since d$z^{k}(u) = 0$ so
\begin{equation*}
DL(u) = (DL)^{\dag}(u) = \textup{d}z^k(u) \frac{\partial L}{\partial \textup{d}z^k} = 0
\end{equation*}
The proper energy density is $T(u,u) = - L$. The proper stress tensor is $P T P = LP - DL$. The proper energy flux $P(T(u)) = 0 = $ proper momentum density. This will not be true when waves are introduced, as things become more complex.

The Euler-Lagrange equations which follow from varying the fields $z^{k}$ (i.e. varying the worldlines of the fluid elements) are
\begin{equation*}
\frac{\partial L}{\partial z^{k}} - \nabla \cdot L_{k} = 0 \quad \textup{where} \quad L_{k} = \frac {\partial L}{\partial z^{k}_{\mu}} e_{\mu}
\end{equation*}
The Lagrangian has no explicit dependence on x, so the Euler-Lagrange equation is equivalent to conservation of energy and momentum:
\begin{equation} \label{eq:divT}
\nabla \cdot T = \nabla_{\nu} T_{\mu}^{\nu} e^{\mu} = 0
\end{equation}
Decomposing equation \eqref{eq:divT} with respect to $u$ gives
\begin{equation}
u \cdot (\nabla \cdot T) = 0 \qquad \textup{and} \qquad P(\nabla \cdot T) = 0
\end{equation}
These are the equations for conservation of energy and momentum.
\section{Perfect Fluids}
The Lagrangian for a perfect fluid is
\begin{equation} \label{eq:lagm}
L = - \rho = - n(1 + e)
\end{equation}
Equations \eqref{eq:nn} and \eqref{eq:defT}  for $n$ and $T$ yield $Dn = nP$ and the well known EMT for a perfect fluid:
\begin{equation} \label{eq:Tfl}
T = Lg - DL = \rho uu + pP
\end{equation}
The proper energy density is $T(u,u) = \rho$. The stress tensor is isotropic; $PTP = pP$, so $T$ is diagonal in a local CMCS; $T = \textup{diag}(\rho , p, p, p)$.
The equations of motion are
\begin{equation*}
\nabla \cdot T = nfa + \textup{d}p + u \nabla \cdot(nfu) = 0
\end{equation*}
where $a = \nabla u(u)$ is the acceleration. 
\subsection{Conservation of Energy}
\begin{equation} \label{eq:fl-enr}
u \cdot (\nabla \cdot T) = \textup{d}p(u) - n \textup{d}f(u) = - nT \textup{d}s(u) = 0
\end{equation}
since $\nabla \cdot (nu) = 0$. This is the thermodynamic equation (1) for adiabatic flow (d$s(u) = 0$). Any two of the following three imply the third:
\begin{align*}
\nabla \cdot (nu) &= 0 & u \cdot (\nabla \cdot T) &= 0 & \textup{d}s(u) &= 0
\end{align*}
For isentropic flow (d$s = 0$), equation \eqref{eq:fl-enr} for the energy is d$p = n \textup{d}f$ .
\subsection{Conservation of Momentum}
\begin{equation} \label{eq:motion}
P(\nabla \cdot T) = nfa + P(\textup{d}p) = 0
\end{equation}
For isentropic flow, d$p = n \textup{d}f$, and the equation of motion \eqref{eq:motion} is
\begin{equation}
a = - P(\textup{d}f)/f
\end{equation}
\section{Magnetohydrodynamics (MHD)}
\subsection{The Electromagnetic Field F}
Decomposing the electromagnetic (EM) field 2-form F with respect to u yields
\begin{equation*}
F = u \wedge E + *(u \wedge B)
\end{equation*}
where $E = - F(u)$ is the proper electric field and $B = *F(u)$ is the proper magnetic field. Both $E$ and $B$ are spatial vector fields, since $F$ is antisymmetric; $E(u) = - F(u,u) = 0$ and $B(u) = *F(u,u) = 0$.
A MHD fluid has infinite conductivity which implies a frozen in magnetic field, i.e. $E = - F(u) = 0$. Thus $F = PFP$ is spatial, and in a global CMCS z,
\begin{equation} \label{eq:emf}
F = *(u \wedge B) = \epsilon_{abc} b^{a} \textup{d}z^{b} \wedge \textup{d}z^{c}/2
\end{equation}
where $\epsilon_{abc}$ is the 3D Levi-Civita tensor and $b^{a}$ are the components of $F$ in the z coordinate system. So we have F in terms of the fields $z^k$. Maxwell's homogeneous equation
\begin{equation*}
\textup{d}F = 0 \quad \Rightarrow \quad \textup{d}b^{a}(u) = 0
\end{equation*}
So the $b^a(z^1, z^2, z^3)$ are constant on streamlines (the field is``frozen in'').
\subsection{The Lagrangian and the Fields}
The Lagrangian for the magnetic field in MHD is $L_{b} = - (F|F)/2 = - B^{2}/2$. Equation \eqref{eq:emf} gives $F$ and thus $L_{b}$ in terms of the fields $z^{i}$ and their first derivatives d$z^{i}$:
\begin{equation}
L_{b} = - \epsilon_{abc} \epsilon_{def} b^{a} b^{d} (\textup{d}z^{b}|\textup{d}z^{e})(\textup{d}z^{c}|\textup{d}z^{f})/4
\end{equation}
The total Lagrangian for MHD is the sum of $L_{b}$ and the Lagrangian for the matter \eqref{eq:lagm} $L_{m} = - \rho$:
\begin{equation}
L = L_{m} + L_{b} =  - \rho - B^{2}/2
\end{equation}
\subsection{The Energy Momentum Tensor (EMT)}
From \eqref{eq:emf} and \eqref{eq:defT} we find that $DL_{b} = BB - B^{2}P$, so the EMT for the magnetic field is
\begin{equation}
T_{b} = L_{b}g - DL_{b} = B^{2}(uu + P) - BB \nonumber
\end{equation}
Using \eqref{eq:Tfl} for $T_m$ gives the total EMT:
\begin{equation} 
\label{eq:emt-mhd}
T = T_{m} + T_{b} = (\rho + B^{2}/2) uu + (p + B^{2}/2)P - BB
\end{equation}
If $e_{1} \parallel B$ in a CMCS, T is diagonal:
\begin{align*}
T^{00} &= \rho + B^{2}/2 & T^{11} &= p - B^{2}/2 & T^{22} &= T^{33} = p + B^{2}/2 
\end{align*}
\subsection{The MHD equations}
The equations for MHD are $\nabla \cdot (nu) = 0$ and
\begin{gather}
\label{eq:mhd-divT}
\nabla \cdot T = \nabla \cdot [(nf + B^2)u]u + (nf + B^2)a + \nabla p + (\nabla B)^{\dag}(B)\\ - (\nabla \cdot B)B - \nabla B(B) = 0 \nonumber \\
dF = 0  \Leftrightarrow  \nabla \cdot *F = \nabla \cdot (Bu - uB) = (\nabla \cdot B)u + \nabla u(B) \nonumber \\
 - (\nabla \cdot u)B - \nabla B(u) = 0
\end{gather}
\subsubsection{Conservation of Energy}
The energy equation for MHD is the projection of equation \eqref{eq:mhd-divT} onto u; $u \cdot (\nabla \cdot T) =0$. It is the same as the energy equation \eqref{eq:fl-enr} for  a perfect fluid.
\subsubsection{Conservation of Momentum}
The MHD equation of motion is the spatial part of \eqref{eq:mhd-divT}:
\begin{equation}
\label{eq:mhd-mom}
P(\nabla \cdot T) = (nf + B^2) a + n P(df) - (\nabla \cdot B)B - P(dB(B)) = 0
\end{equation}
Maxwell's inhomogeneous equation relates the electric current 4-vector $K$ to $B = *F(u)$:
\begin{equation*}
K = *d*F = *d(B \wedge u) = *dB(u) - *du(B) .
\end{equation*}
The MHD equation of motion can be written as
\begin{equation*}
P(\nabla \cdot T) = (nf + B^2) a + n P(df) - K \times B = 0
\end{equation*}
where is Lorentz force is $K \times B = P(K) \times B = *(u \wedge K \wedge B)$.

\part{Waves in Continua}
\section{Formulating the Problem}
The following theory of waves in continua gives the dispersion relation and polarization of the various modes, the bi-characteristic rays, the equation governing the wave amplitude, the ray equations which determine the wave vector
along the rays, and the EMT of the waves, the average background flow, and interaction between the waves and the average background flow. There are three 4-vector fields associated with any wave mode. They are
\begin{enumerate}
\item The wave 4-vector $k = \omega u + \kappa$, where $\omega = -k(u)$ and $\kappa = P(k)$ are the proper frequency and wavenumber vector.
\item The phase 4-velocity $v_\phi = u + v_p$ is the 4-velocity of a surface of constant phase. $P(v_\phi) = v_p = \omega \kappa/\kappa^2$ is the (proper) spatial phase velocity with respect to the fluid, i.e. as seen in a CMCS.
\item The group 4-velocity $g = u + v_g$ is the 4-velocity of wavefronts (or of wave quanta or wave packets). $P(g) = v_g = \partial \omega/\partial \kappa$ is the spatial group velocity in a CMCS.
\end{enumerate}
Waves in continua are described by the \emph{characteristic} hypersurfaces (with normal k) and bi-characteristic rays (integral curves of w). k satisfies the dispersion relation. Wavefronts can be thought of as discontinuities in the solutions to the equations of motion. Consider plane waves with surfaces of constant phase given by $\varphi (x) = (k|x) = k_\mu x^\mu$, so $k = d\varphi$ is the normal to the hypersurfaces $\varphi(x) =$ constant.
\subsection{The Two Sets of Independent Coordinates and Fields}
Make the usual assumption that the average background flow varies little over a period or a wavelength of the waves (the WKB approximation). The following theory of the waves and their interaction with the background flow requires two sets of independent coordinates and fields, i.e. two maps from Euclidean to Lagrangian coordinates, together with a Lorentz transformation from the local CMCS y to the x coordinate system (CS) of the observer. The Lagrangian is a scalar field so it can be calculated using the coordinates that are most convenient. The two sets of coordinates and fields must be clearly defined. They are:
\begin{enumerate}
\item $z^i \colon x \to z^i(x)$ ;  $(x,z)$ refers to the average flow ;
\item $w^i \colon y \to w^i(y) = y^i + \phi^i(y)$  where  $\phi^i(y) = \alpha^i \cos \varphi(y)$ ;  
\end{enumerate}

$(x,z)$ defines the slowly varying flow of the average position of the fluid elements (i.e. the flow of the oscillation centers), called the average or background flow. $x$ is the position of an oscillation center, rather than a fluid element. $z^i(x)$ is the initial position of the oscillation center of the fluid element whose oscillation center is at $x$.

$(y,w)$ defines the waves in a local CMCS $y$ in the neighborhood of any $x \in M$. The $y$ CS is moving with the 4-velocity u of the fluid's oscillation centers, so that $e_0 = \partial/\partial y^0 = u$ in this CS. The fields $w^i(y)$ are the oscillation centers, the average position of the fluid element at $y$. $\phi^i(y)= \alpha^i \cos \varphi(y)$ is the displacement due to waves. The spatial vector $\alpha = (\alpha^1, \alpha^2, \alpha^3)$ is the wave amplitude. For a plane wave, the phase $\varphi(y) = (k|y)$ is rapidly varying, but $\alpha$ and $k = d\varphi$ are slowly varying functions of $x$ (not $y$). To treat the waves, we replace the fields $w^i$ by $\alpha^i$ and $\varphi$, the appropriate fields for waves. Varying $\alpha$ gives the dispersion relation and polarization $\hat\alpha = \alpha/|\alpha|$. Varying $\varphi$ gives the group velocity, the rays, and the equation for conservation of wave action (or wave quanta--dividing by Planck's constant gives a dimensionless number that is conserved, which I call quanta, even though this is not a quantum theory). This determines how the wave amplitude evolves as the wave propagates through the fluid.
\section{The Average Lagrangian}
In the following, I will denote rapidly varying quantities with a prime, and averaged ones as unprimed.

The (slowly varying) average Lagrangian $L(z^k, dz^k, \alpha, k; x)$ of the total wave-background system is derived as follows. Make a Taylor expansion about the background flow (of the oscillation centers) of the rapidly varying Lagrangian $L'$ of the actual flow, to order $\alpha^2$. Then average $L'$ over the phase $\varphi$ (Whitham 1974) to get the total average Lagrangian $L = L_0 + L_w$, where $L_0$ is the Lagrangian of the background and $L_w$ is the wave Lagrangian. One can argue that the rapidly varying part of L would average out in the integration in Hamilton's principle anyway.
\subsection{Sound Waves in a Perfect Fluid}
Apply the coordinates $(y,w)$ to \eqref{eq:nn}. We have
\begin{gather}
\label{eq:nnn}
n'(y) = r(w) \sqrt{det(dw^i|dw^j)} = n + \delta n\\
dw^i = dy^i + d\phi^i = e^i - \alpha^i k \sin \varphi \nonumber
\end{gather}
To order $\alpha^2$ we have
\begin{equation}
\label{eq:nnw }
\delta n/n = - \kappa \cdot \alpha \sin \varphi - \omega^2 \alpha^2 \sin^2 \varphi/2
\end{equation}
Expand $L' = - \rho' = - n'[1 + e(n')] $about $n'$ to order $\alpha^2$:
\begin{equation*}
L' = L_0(n + \delta n) = L_0(n) + \frac{\partial L_0}{\partial n} \delta n + \frac{\partial^2 L_0}{\partial n^2} (\delta n)^2/2 + \cdots
\end{equation*}
where $L_0(n) = - \rho(n) = - n(1 + e)$ is the Lagrangian of the background,
\begin{equation*}
\frac{\partial L_0}{\partial n} = - f, \quad \textup{and} \quad \frac{\partial^2 L_0}{\partial n^2} = - \frac{fc^2}{n}  ; \quad \textup{where} \quad c^2 = \frac{dp}{d\rho} = \frac{\gamma T}{mf}
\end{equation*}
c is the speed of sound.

The average of any function $f(\varphi)$ is
\begin{gather*}
\langle f(\varphi) \rangle = \frac{1}{2\pi} \int_0^{2\pi}{f(\varphi)}d\varphi \quad \Rightarrow \quad \langle \phi^i \rangle = 0 ,\quad \langle w^i \rangle = y^i \\
 \langle \delta n \rangle = - n \omega^2 \alpha^2/4 ,\qquad  \langle( \delta n)^2 \rangle = n^2 (\kappa \cdot \alpha)^2/2
\end{gather*}
The total average Lagrangian is 
\begin{equation}
\label{eq:lag1}
L = \langle L' \rangle = L_0 + L_w = - n(1 + e) + nf[\omega^2 \alpha^2 - c^2(\kappa \cdot \alpha)^2]/4
\end{equation}
For ideal gases, relativistic kinetic theory gives exact expressions for $n, e, \rho, p, \gamma$, and $c$ as functions of T/m (Lightmann et.al. 1975). For a monatomic ideal gas, as $T \to 0$, $\gamma \to 5/3$ and $c^2 \to 5T/3m$, and as $T \to\infty, \rho \to 3p$, so $c^2 = dp/d\rho \to 1/3$ and $\gamma \to 4/3$.
\subsection{Magnetohydrodynamic (MHD) Waves}
Repeating this procedure for MHD, expand $L' = L'_m + L'_b = - \rho' - (F'|F')/2$ to order $\alpha^2$ and average over $\varphi$ to get 
\begin{equation}
\label{eq:lagmhdd}
L = \langle L' \rangle = L_{0m} + L_{0b} + L_{wm} + L_{wb}
\end{equation}
The subscript {\footnotesize 0} refers to the background flow, w refers to the waves, m to the matter, and b to the magnetic field.
 
Equation \eqref{eq:lag1} gives $\langle L'_m \rangle = L_m = L_{0m} + L_{wm}$. The calculation for $L_b$ is analogous to that for \eqref{eq:nnn}. Using equation \eqref{eq:emf} for $F' = *(u \wedge B')$ and substituting $(y,w)$ for $(x,z)$ in $L'_b$ gives
\begin{equation}
\label{eq:lagmhd1}
L'_b = - \epsilon_{abc} \epsilon_{def} B^a B^d (dw^b|dw^e)(dw^c|dw^f)/4
\end{equation}
Putting $dw^i =  e^i - \alpha^i k \sin \varphi$ in \eqref{eq:lagmhd1}  and averaging yields
\begin{gather}
\label{eq:lagmhd2}
L_b = \langle L'_b \rangle = - B^2/2 + L_{bw} , \quad \textup{where} \nonumber \\
4L_{bw} = k^2[(\alpha \cdot B)^2 - \alpha^2 B^2] + (B \wedge \alpha \wedge \kappa|B \wedge \alpha \wedge \kappa)
\end{gather}
is the part of $L_w$ due to the magnetic field. After averaging, set $\langle w^i \rangle = y^i \to x^i$ since  $x^i$ is the position of the oscillation center of the fluid element. Thus the total average Lagrangian \eqref{eq:lagmhdd} is
\begin{gather}
L = L_0 + L_w \; ; \; L_0 = - \rho - B^2/2 \quad \textup{and} \quad L_w = L_{wm} + L_{wb}\nonumber \\
\label{eq:lagmhd3}
4L_w = \omega^2\alpha^2[nf + B^2 - (\hat \alpha \cdot B)^2] - (nfc^2 + B^2)(\kappa \cdot \alpha)^2\\
 - \alpha^2(\kappa \cdot B)^2 + 2(\kappa \cdot \alpha)(\kappa \cdot B)(\alpha \cdot B) \nonumber
\end{gather}
\subsubsection{Alfven Waves: $B \cdot \alpha = \kappa \cdot \alpha = 0$}
For Alfven waves \eqref{eq:lagmhd3} gives
\begin{equation}
\label{eq;lagal}
2 L_w = (nf + B^2)\omega^2\alpha^2 - (\kappa \cdot B)^2\alpha^2
\end{equation}
\subsubsection{Magnetosonic Waves: $B \wedge \alpha \wedge \kappa = 0$}
For the fast and slow magnetosonic modes  \eqref{eq:lagmhd3} gives
\begin{equation}
\label{eq:lagms}
4L_w = [nf + B^2 - (\hat\alpha \cdot B)^2]\omega^2 \alpha^2 - nfc^2(\kappa \cdot \alpha)^2 - \kappa^2 \alpha^2 [B^2 - (\hat\alpha \cdot B)^2]
\end{equation}

\section{The Dispersion Relation, Phase Velocity, and Polarization}
To discuss the dispersion and polarization relations write $L_w$ as a symmetric quadratic function of the wave amplitude $\alpha$ on the 3D space orthogonal to u (at each $x \in M$):
\begin{equation*}
L_w = C(\alpha,\alpha)/2
\end{equation*}
where $C$ is a 3 x 3 matrix. The fields for the waves are $\alpha$ and $\varphi$,
the action-angle coordinates. The Euler-Lagrange equation for $\alpha$ is
\begin{equation*}
\frac{\partial L_w}{\partial \alpha} = C(\alpha) = 0 \quad \Rightarrow \quad \det C = 0
\end{equation*}
which is the dispersion relation. $\det C = 0$ is a cubic for $v^2_p = \omega^2/\kappa^2$, so there are 3 solutions for $v^2_p$ and 3 corresponding solutions of $C(\hat \alpha) = 0$ for the polarization $\hat \alpha = \alpha/|\alpha|$. The phase 4-velocity is $v_{\phi} = u \pm v_p$.
\subsection{Sound Waves}
Writing equation \eqref{eq:lag1} in the form $L_w = C(\alpha,\alpha)/2$ yields
\begin{equation*}
C = nf (\omega^2 P - c^2 \kappa \kappa)
\end{equation*}
P is the identity matrix of the 3D space orthogonal to u. If $e_1\parallel \kappa$,  $C$ is diagonal, the dispersion relation is $\det C = \omega^4(\omega^2 - c^2 \kappa^2) = 0$. $v_p = 0$ when $\kappa \cdot \alpha = 0$, corresponding to transverse modes which are comoving discontinuities and do not propagate. The longitudinal mode with $\kappa \parallel \alpha$ is a sound wave with
\begin{equation}
\omega^2 = c^2 \kappa^2 \quad \Rightarrow \quad v_p = v_g = c \hat\kappa
\end{equation}
\subsection{MHD Waves}
Let $e_1 \parallel B$, $\kappa$ in the $(e_1, e_2)$ plane, and $\cos \theta = \hat \kappa \cdot \hat B$. Then
\[
\mathbf{C} =
\begin{pmatrix}
C_{11} & C_{12} &0 \\
C_{21} & C_{22} & 0 \\
0 & 0 & C_{33}
\end{pmatrix}
\]
and $\det C = C_{33}[C_{11}C_{22} - (C_{12})^2]$. The $C_{ij}$ are given below. MHD waves fall into two classes, depending on the polarization $\hat \alpha$. When $\alpha^1 = \alpha^2 = 0$, $\hat \alpha = e_3$ is $\perp$ to the plane containing $B$ and $\kappa$. These are Alfven waves with dispersion relation $C_{33} = 0$.When $\alpha^3 = 0$ and $\alpha$ lies in the plane of $B$ and $\kappa$, we have magnetosonic waves with dispersion relation
\begin{equation*}
C_{11}C_{22} - (C_{12})^2 = 0
\end{equation*}
\subsubsection{Alfven Waves}
Alfven waves are transverse, with $\alpha \perp$ both $\kappa$ and $B$, and equation \eqref{eq;lagal} for $L_w = C_{33}\alpha^2/2$ gives the dispersion relation
\begin{gather}
\label{eq:alf1}
v_p^2 = \frac{(B \cdot \hat \kappa)^2}{nf + B^2} = v_A^2 \cos^2 \theta \quad; \quad v_p = v_A \cos \theta \: \hat \kappa \quad \textup{where} \quad v_A^2 = \frac{B^2}{nf + B^2}
\end{gather}
\subsubsection{Magnetosonic Waves}
$B, \kappa$, and $\alpha$ all lie in the same plane, so $B \wedge \kappa \wedge \alpha = 0$ (see \eqref{eq:lagms} for $L_w$). The dispersion relation for the fast ($+$) and slow($-$) (compressive) magnetosonic modes is $\det C = 0$, which is a quadratic equation for the square of the phase velocity $v^2_{\pm} = \omega^2/\kappa^2$ :
\begin{align}
\label{eq:msdr}
2C &= (nf + B^2)\omega^2 P - \omega^2 BB - nfc^2\kappa \kappa - \kappa^2(B^2P - BB) \quad \Rightarrow \nonumber \\
2v_{\pm}^2 &= c^2 + v_A^2 - c^2 v_A^2 \sin^2\theta \pm R \quad \textup{where}\\
R^2 &= (c^2 - v_A^2)^2 + 2 c^2 v_A^2\sin^2\theta[2 - c^2 - v_A^2 + c^2 v_A^2\sin^2\theta/2] \quad \textup{and} \nonumber \\
v_p & = v_{\pm}\: \hat \kappa \nonumber
\end{align}
Note that $v_{-} \le v_+$ and $c$ and $v_A$ appear symmetrically in $v_{\pm}$.
\subsection{Polarization}
The polarization of sound waves ($\alpha \parallel \kappa$) and Alfven waves ($\alpha \perp \kappa$) is clear. For the magnetosonic modes, let $\cos \beta = \hat \alpha \cdot \hat B$. Then $\hat \alpha_{\pm} = \cos \beta e_1 + \sin \beta e_2$, and $C(\hat \alpha) = 0 \Rightarrow$
\begin{equation}
\label{eq:pol}
\tan \beta = - \frac{C_{11}}{C_{12}} = - \frac{C_{12}}{C_{22}}
\end{equation}
The Lagrangian \eqref{eq:lagms} and the dispersion relation \eqref{eq:msdr} give the components of C as functions of $\theta$, the angle  between $B$ and $\kappa$. Then \eqref{eq:pol} gives the polarization $\beta(\theta)$ and $\hat \alpha(\theta)$. The waves are discussed in detail in another paper (reference here).
\section{Wave Quanta, Group Velocity, and Rays}
Wave action can be replaced by the equivalent, more intuitive concept of wave quanta (let Planck's constant  $\hbar$ = 1). Even though everything is continuous, we will refer to the wave action as \emph{wave quanta}. The discussion of wave quanta, group velocity, and the rays, is clarified by writing the wave Lagrangian as a symmetric function of k:
\begin{equation}
L_w = Q(k,k)/2
\end{equation}
where $Q = Q^\dag$ is symmetric. The Euler-Lagrange equation for $\varphi$ expresses conservation of wave quanta. Since $\varphi$ has been averaged over, we have
\begin{equation}
\label{eq:act}
\frac{\partial L_w}{\partial \varphi} - \frac{\partial}{\partial x^\mu} \frac{\partial L_w}{\partial \partial_\mu\varphi} = - \nabla \cdot \frac{\partial L_w}{\partial k} = \nabla \cdot A = 0
\end{equation}
where the wave quanta current 4-vector
\begin{equation}
\label{eq:aa}
A \equiv - \frac{\partial L_w}{\partial k} = - Q(k) = Nu + P(A) = N(u + v_g) = Ng
\end{equation}
is a conserved current,
\begin{equation}
N = - A(u) = \frac{\partial L_w}{\partial \omega} \quad \textup{and} \quad P(A) = Nv_g = - \frac{\partial L_w}{\partial \kappa}
\end{equation}
are the proper density and flux of wave quanta. The (proper) wave energy density E and group velocity are
\begin{align}
\label{eq:Evg }
E &= N \omega & v_g = \frac{\partial \omega}{\partial \kappa} = \frac{P(A)}{N} = - \frac{\partial L_w/\partial \kappa}{\partial L_w/\partial \omega}
\end{align}
Equation \eqref{eq:aa} for $A$ shows that the wave quanta move with the group velocity. 

$\Sigma$ is the hypersurface defined by $x^0 = t = 0$, with normal $dt = e^0$ and volume element $*(e_0) = e^{123} = dx^1 \wedge dx^2 \wedge dx^3 = d^3x$. The conserved quantity associated with the conserved current $A$ is the total number of wave quanta $N_w$:
\begin{equation*}
N_w = \int_{\Sigma}{*A} = \int_{\Sigma}{A^0d^3x}
\end{equation*}
\subsection{Sound Waves}
Writing equation\eqref{eq:lag1} as $L_w = Q(k,k)/2$ shows that
\begin{align*}
Q &= nf(\alpha^2 uu  -c^2 \alpha \alpha)/2\\
A &= - Q(k) = N(u + v_g) = nf\alpha^2(\omega u + c^2 \kappa)\\ 
N &= nf\omega \alpha^2/2 = E/\omega \\
v_g &= c \hat \kappa\; ; \quad g = u \pm c \hat \kappa
\end{align*}
\subsection{MHD Waves}
\subsubsection{Alfven Waves}
Equation \eqref{eq;lagal} for $L_w$ gives
\begin{align*}
2Q &= (nf + B^2) \alpha^2 uu - \alpha^2 BB \\
2A &= - 2Q(k) = (nf + B^2) \alpha^2 \omega u + \alpha^2 (\kappa \cdot B)B \\
N &= - A(u) = (nf + B^2) \omega \alpha^2/2 = E/\omega\\
P(A) &= Nv_g = \alpha^2 (\kappa \cdot B)B/2 \\
\Rightarrow \quad v_g &= v_A \hat B \quad \textup{and} \quad g = u \pm v_A \hat B
\end{align*}
\subsubsection{Magnetosonic Waves}
Define $\cos \eta = \hat \kappa \cdot \hat \alpha$, and recall $\cos \theta = \kappa \cdot \hat B$ and $\cos \beta = \hat \alpha \cdot \hat B$, so $\theta = \beta + \eta$, since the 3 vectors all lie in one plane. Equation \eqref{eq:lagms} for $L_w$ gives
\begin{align*}
2Q  & = (nf + B^2 \sin^2\beta)\alpha^2 uu - nfc^2\alpha\alpha - B^2\alpha^2\sin^2\beta P \\
A  & = - Q(k) = Nu + P(A)\\
N & = - A(u) = \omega\alpha^2 (nf + B^2 \sin^2\beta)/2 = E/\omega\\
2P(A) & = 2Nv_g = nfc^2(\kappa \cdot \alpha) \alpha + B^2 \alpha^2 \sin^2\beta \kappa \\
\Rightarrow \quad v_g^\pm(\theta) & = \frac{(1 - v_A^2)c^2\cos \eta_{\pm} \hat \alpha_{\pm} + v_A^2 \sin^2 \beta_{\pm} \hat \kappa}{v_{\pm}(1 - v_A^2 \cos^2 \beta_{\pm})}
\end{align*}
$v_g^-$ is confined to a narrow cone whose axis is $B$, but $v_g^+$ can be in any direction. $v_g^{\pm}$, like $v_{\pm}$, is unchanged when $c^2$ and $v_A^2$ are interchanged. Also 
\begin{gather*}
2L_w = Q(k,k) = k \cdot A = N k \cdot g= 0 \\
v_p^2 = v_g \cdot v_p \Rightarrow v_p \le v_g
\end{gather*}
\subsection{The Normal Cone, The Ray Cone, and the Ray Equations}
The bi-characteristic rays are the world lines of wave quanta and integral curves of the group 4-velocity $g = u \pm v_g$ or the wave quanta current $A = Ng$. There are two directions for the rays of each mode, corresponding to $g = u + v_g$ and $g = u - v_g$. The 3D characteristic hypersurfaces \{$C: \varphi(x) =$ constant\} are wavefronts, across which discontinuities occur.  Since k = $d\varphi$ and $k \cdot g = 0$, rays lie in characteristic surfaces. The wavefront for a disturbance at the origin $x = 0$ divides spacetime M into two regions: behind or ahead of the wavefront. The wavefront can be thought of as the envelope of the surfaces orthogonal to the rays formed by a burst of wave quanta from the origin.

Q defines two quadratic cones at each point of spacetime: the \emph{normal} cone $Q(k,k) = 0$, which is generated by the rays which are integral curves of $A$, and the \emph{ray} cone $Q^{-1}(A,A) = 0$, dual to the normal cone, which is generated by the wave-vector k. Each of the two cones is the envelope of the planes orthogonal to the rays of the other. It is useful to look at the spatial cross sections of these cones obtained by setting the time components equal to $1$. In MHD, the surfaces have axial symmetry about the magnetic field. The cross section of the normal cone $Q(k,k) = 0$ is called the normal or wave-vector surface. The cross section of the ray cone $Q^{-1}(A,A) = 0$ is called the ray surface. Its useful to plot $v_g(\theta)$ and $v_p(\theta)$. For a further discussion of the non-relativistic results see Courant and Hilbert Vol. II, \S VI (1962).

Since $k = d \varphi$, $dk = d^2\varphi = 0$. If $k'_\mu = \partial \varphi/\partial x^\mu$ are the components of k in the x coordinate system (CS),
\begin{gather}
\label{eq:dk}
dk = \left(\frac{\partial k'_\nu}{\partial x^\mu} - \frac{\partial k'_\mu}{\partial x^\nu}\right) dx^\mu \wedge dx^\nu/2 = 0 \\
\Rightarrow \quad \frac{\partial \kappa'_i}{\partial t} = - \frac{\partial \omega'}{\partial x^i} \quad \textup{and} \quad \frac{\partial \kappa'_i}{\partial x^j} = \frac{\partial \kappa'_j}{\partial x^i} \nonumber
\end{gather}
where $\omega' = \omega'(\kappa'(x),x)$ is the dispersion relation. $k'_\mu = ( - \omega', \kappa'_i)$ are related to the proper components of $k_\mu = ( - \omega, \kappa_i)$ by a Lorentz transformation with the local fluid velocity $v(x)$:
\begin{equation}
\omega' = \gamma(\omega + \kappa \cdot v) \quad ; \quad \kappa'_\parallel = \gamma( \kappa_\parallel + \omega v) \quad ; \quad \textup{and} \quad \kappa'_\perp = \kappa_\perp
\end{equation}

Consider the dispersion relation as a function of x, through both k and the inhomogeneities in the background flow:
\begin{align*}
dL_w(k(x), x)  & = \left[ \frac{\partial L_w}{\partial k_\mu} \frac{\partial k_\mu}{\partial x^\nu} + \frac{\partial L_w}{\partial x^\nu} \right] dx^\nu =  - \nabla k(A) + \left[ \frac{\partial L_w}{\partial x} \right]_k = 0
\end{align*}
where $\nabla k(A) \equiv \nabla_A k = A^\mu \nabla_\mu k$. Now let $L_w(k,x)$ be the dispersion relation on the 8-dimensional phase space $(k,x)$, with both $k$ and $x$ as independent coordinates. The ray equations are
\begin{equation}
\label{eq:ray}
\frac{dk}{d\sigma} = \nabla k(A) = \left[\frac{\partial L_w}{\partial x}\right]_k \quad \textup{and} \quad \frac{dx}{d\sigma} = \nabla x(A) = A = - \left[ \frac{\partial L_w}{\partial k}\right]_x
\end{equation}
where $\sigma$ is a parameter along the ray. Equation \eqref{eq:ray} is the 4D canonical Hamiltonian form of the ray equations for the waves. Since $A = Ng$ we can divide the ray equations by $N = \partial L_w/\partial \omega \Rightarrow$
\begin{equation*}
\nabla k(g) = \frac{1}{N} \frac{\partial L_w}{\partial x} = - \left[\frac{\partial \omega}{\partial x}\right]_\kappa \quad \textup{and} \quad \nabla x(g) = g = - \frac{1}{N} \frac{\partial L_w}{\partial k} = \frac{A}{N} = u + \frac{\partial \omega}{\partial \kappa}
\end{equation*}
Its always useful to write the equations in coordinate free form, since then they can be evaluated in any CS.
\section{The Energy Momentum Tensor (EMT)}
The total canonical EMT for a system consisting of a slowing varying background flow interacting with waves is
\begin{equation*}
T = Lg - DL - k \frac{\partial L}{\partial k} \quad \textup{where} \quad L = L_0(z^i, dz^i; x) + L_w(z^i, dz^i, \alpha, k; x)
\end{equation*}
The EMT T divides naturally into 3 parts:
\begin{align}
\label{eq:emt}
T  &= T_0 + I + W =  \textup{background} + \textup{interaction} + \textup{waves,} \quad \textup{where}\\
T_0  &= L_0 g - DL_0 \quad ; \quad I = - DL_w \quad ; \quad W = - k \frac{\partial L_w}{\partial k} = k A \nonumber 
\end{align}
$T_0$ is the EMT of the background alone, since $D = dz^k \otimes \partial /\partial dz^k$ and $L_0$ depend only on the background fields. Since I is the derivative of $L_w$ with respect to the background fields, it is the EMT of the wave-background interaction. W involves only the wave field $k$ and $L_w$, so it is the EMT of the waves alone.

We need to know the action of D on all the quantities in L:
\begin{align}
\label{eq:dd}
Dn &= nP   & DJ & = Jg - gJ\\
Du &= - Pu & D\omega & = \kappa u \nonumber \\
D\kappa &= \omega Pu - u\kappa u & Dk & = D(\omega u + \kappa) = 0 \nonumber \\
D(\kappa\cdot \alpha) &= \omega \alpha u & DT &= (\gamma - 1)TP &  \nonumber \\
DB &= BP - PB - uBu & DB^2 &= 2(B^2P - BB) \nonumber\\
D(B \cdot \kappa) &= (B \cdot \kappa)P - \kappa B + \omega Bu & D(B \cdot \alpha) & = (B \cdot \alpha)P - \alpha B \nonumber \\
Df & = fc^2 P & D\gamma &= - \zeta \gamma P \nonumber \\
Dc^2 &= 2 \gamma' c^2P & \zeta &= \frac{T}{c_p c_v^2}\frac{dc_v}{dT} \nonumber
\end{align}
and $\gamma' = (\gamma - 1 - c^2 - \zeta)/2$.
\subsection{The Proper Components of the EMT}
The decomposition of the parts of the EMT with respect to u yields the proper energy and momentum density and flux. The wave EMT $W$ is
\begin{equation}
\label{eq:emtw}
W = k \otimes A = N(\omega u + \kappa)\otimes(u + v_g) = Euu + Euv_g + N\kappa u + N \kappa v_g
\end{equation}
The total EMT due to the waves is $T_w = I + W =$ the wave $+$ interaction EMT. Since L depends only on scalar fields, $T$, $T_0$, and $T_w$ are symmetric, but I and W are not. We have $T = T_0 + T_w$, where $T_0$ is the EMT of the background given in Part 1, and
\begin{align}
\label{eq:emtw1}
T_w &= I + W = E(uu + uv_g + v_g u) + N \kappa v_g + PIP 
\end{align}
 \subsubsection{The Components of the EMT for Waves}
The energy and momentum density and flux for the waves are the components of $W$ and $T_w$. They can be read off from equations \eqref{eq:emtw} and \eqref{eq:emtw1}. The wave energy density is
\begin{equation*}
W(u,u) = T_w(u,u) = E = N\omega
\end{equation*}
since the interaction energy density $I(u,u) = 0$. The interaction energy flux density is $P(I(u)) = 0$ also. The wave momentum density is $- P(W^\dag(u)) = N \kappa$, as one would expect. However, this term in W is canceled by one in I to give the total wave + interaction momentum density
\begin{equation*}
- P(T^\dag_w(u)) = - P(I^\dag(u) + W^\dag(u)) = Ev_g = - P(T_w(u))
\end{equation*}
as one would expect. The wave pressure tensor is
\begin{equation*}
PT_wP = PIP + PWP = PIP + N\kappa v_g
\end{equation*}
which is always symmetric.
\subsection{Sound Waves}
The EMT of the background is $T_0 = \rho uu +pP$. For the waves, $v_p = v_g = c\hat\kappa$, and
\begin{equation*}
T_w = I + W = E(uu + cu\hat\kappa + c\hat\kappa u + \hat\kappa\hat\kappa + \gamma' P)
\end{equation*}
The components of $T_w$ in a CMCS are
\begin{align*}
T_w(u,u) &= E = N\omega & - P(T_w(u)) = - P(T^\dag_w(u)) = E c \hat\kappa \\
PT_wP& = E(\gamma' P + \hat\kappa\hat\kappa)
\end{align*}
The components of the total EMT  are
\begin{align}
\label{eq:emts}
 T = T_0 +T_w =  (\rho + E) uu + Ec(u \hat\kappa + \hat\kappa u) + E\hat\kappa\hat\kappa + (p + \gamma')P
\end{align}
\subsection{MHD Waves}
The EMT of the background for MHD was given in equation \eqref{eq:emt-mhd}. 
\subsubsection{Alfven Waves}
The phase and group velocities are $v_p = v_A \cos \theta \hat \kappa$ and $v_g = v_A \hat B$ (see eq. \eqref{eq:alf1} and \S10.2.1. Using \eqref{eq:dd} to calculate the EMT yields
\begin{align*}
T_w  &= E(uu + v_A u \hat B + v_A \hat B u) + \omega^2 \alpha^2[nf(1 - c^2)P + 2BB]/4 \\
 T &= T_0 + T_w  
\end{align*}
\subsubsection{Magnetosonic Waves}
The phase velocity, group velocity and energy density of the fast and slow magnetosonic modes are given in \S9.2.2 and \S10.2.2. The EMT for the waves is
\begin{align*}
  T_w  &= E(uu + uv_g + v_g u) + PT_wP 
\end{align*}
The expression for the wave pressure tensor $PT_wP$ is quite complex and I will not reproduce it here.
\subsection{The Equations for The Total Wave-Background System}
The following equations completely determine the waves and the background flow:
\begin{align*}
 \nabla \cdot T  &= 0 & \nabla \cdot J &= 0 & \nabla \cdot A &= 0   
\end{align*}
and the ray equations \eqref{eq:dk} and \eqref{eq:ray}. One \emph{must} include the EMT of the waves $T_w$ when it is comparable to $T_0$, i.e., when the wave energy density E is comparable to the thermal or magnetic field energy density $p$ or $B^2$, e.g. in the solar wind. Waves transport energy and momentum in ways in which other processes cannot. A similar treatment can be given for plasma waves (Jacques 1988) and wave in other continua.


\begin{thebibliography}{99}

\bibitem{bre} Bretherton, F P (1970) in {\it Mathematical Problems in the Geophysical Sciences}\,, Providence: American Mathematical Society.

\bibitem{AMP} Choquet-Bruhat, Y, DeWitt-Morette, C \& Dillard-Bleick, M (1977) {\it Analysis, Manifolds and Physics}\,, New York: North Holland.

\bibitem{CH} Courant, R and Hilbert, D (1962) {\it Methods of Mathematical Physics II}\,,  New York: Wiley-Interscience.

\bibitem{dew} Dewar, R L (1970) {\it Physics of Fluids}\, {\bf 13}\,, 2710.

\bibitem{lich} Lichnerowicz, A (1967) {\it Ann. Inst. Poincare}\, {\bf 7}\,, 271.

\bibitem{light} Lightmann, et. al. (1975) {\it A problem Book in Relativity and Gravitation}\,, Princeton University Press.

\bibitem{mtw} Misner, C W , Thorne, K S and Wheeler, J A (1973)  {\it Gravitation}\,, San Francisco: Freeman.

\bibitem{sop} Soper, D E (1976) {\it Classical Field Theory}\,, New York: Wiley-Interscience.

\bibitem{thir} Thirring, W A (1980) {\it A Course in Mathematical Physics II: Classical Field Theory}\,, New York: Springer-Verlag

\bibitem{whit} Whitham, G B (1974) {\it Linear and Nonlinear Waves}\,, New York: Wiley-Interscience.

\end{thebibliography}
\end{document}